\documentclass[aps,graphicx,6pt,showkeys,showpacs]{revtex4}
\usepackage{amsmath}
\usepackage{amscd}
\usepackage{graphicx}
\usepackage{subfigure}
\usepackage{appendix}

\begin{document}

\title[Short Title]{Reverse engineering of a Hamiltonian by designing the evolution operators}

\author{Yi-Hao Kang$^{1}$}
\author{Ye-Hong Chen$^{1}$}
\author{Qi-Cheng Wu$^{1}$}
\author{Bi-Hua Huang$^{1}$}
\author{Yan Xia$^{1,}$\footnote{E-mail: xia-208@163.com}}
\author{Jie Song$^{2}$}

\affiliation{$^{1}$Department of Physics, Fuzhou University, Fuzhou 350002, China\\
             $^{2}$Department of Physics, Harbin Institute of Technology, Harbin 150001, China}

\begin{abstract}
We propose an effective and flexible scheme for reverse engineering
of a Hamiltonian by designing the evolution operators to eliminate
the terms of Hamiltonian which are hard to be realized in practice.
Different from transitionless quantum driving (TQD) \cite{sa3}, the
present scheme is focus on only one or parts of moving states in a
$D$-dimension ($D\geq3$) system. The numerical simulation shows that
the present scheme not only contains the results of TQD, but also
has more free parameters, which make this scheme more flexible.
An example is given by using this scheme to realize the population
transfer for a Rydberg atom. The influences of various decoherence
processes are discussed by numerical simulation and the result shows
that the scheme is fast and robust against the decoherence and
operational imperfection. Therefore, this scheme may be used to
construct a Hamiltonian which can be realized in experiments.

\pacs{03.67. Pp, 03.67. Mn, 03.67. HK}

\keywords{Reverse engineering; Shortcuts to adiabatic passage;
Transitionless quantum driving}
\end{abstract}

\maketitle

\section*{Introduction}

Executing computation and communication tasks \cite{ex1,ex2,ex3,ex4}
with time-dependent interactions in quantum information processing
(QIP) \cite{sheng1,sheng2,sheng3,sheng4,an1,an2,an3,an4} have
attracted more and more interests in recent years. It has been shown
that, the adiabatic passage, resonant pulses, and some other methods
can be used to realize the evolution process. Among of them, the
adiabatic passage techniques are known for their robustness against
variations of experimental parameters. Therefore, many schemes have
been proposed with adiabatic passage techniques in quantum
information processing field. For example, rapid adiabatic passage,
stimulated Raman adiabatic passage, and their variants
\cite{ad1,ad2,ad3,ad4,ad5,ad6,ymd5,ymd6,ymd8,ymd9} have been widely
used to perform population transfers in two- or three-level systems.
The system keeps in the instantaneous ground state of its
time-dependent Hamiltonian during the entire evolution process under
an adiabatic control of a quantum system. To ensure that the
adiabatic condition is always satisfied, the control parameters in
the Hamiltonian should be well designed, which usually issue in
relatively long execution time. Although little heating or friction
will be created when the system remains in the instantaneous ground
state, the long time required may make the operation useless or even
impossible to implement because decoherence would spoil the intended
dynamics. On the other hand, using resonant pulses, the scheme may
has a relatively high speed, but it requires exact pulse areas and
resonances. Therefore, accelerating the adiabatic passage towards
the perfect final outcome is a good idea and perhaps the most
reasonable way to actually fight against the decoherence that is
accumulated during a long operation time. Consequently, some
alternative approaches have been put forward by combining the
virtues of adiabatic techniques and resonant pulses together for
achieving controlled quantum state evolutions with both high speed
and fidelity, such as optimal control theory \cite{oc1,oc2,oc3} and
composite pulses \cite{cp1,cp2}. Recently, by designing nonadiabatic
shortcuts to speed up quantum adiabatic process, a new technique
named ``shortcuts to adiabaticity'' (STA)
\cite{sa1,sa2,sa3,sa9,sa4,sa5,sa6,sa7,sa8,c2,ymd4,addd1} opens a new
chapter in the fast and robust quantum state control. As two famous
methods of STA, ``Transitionless quantum driving'' (TQD)
\cite{sa1,sa3,sa9,sa4} and inverse engineering
\cite{c2,sa4,sa7,sa8,addd1} based on Lewis-Riesenfeld invariants
\cite{lr} have been intensively focused, They have been applied in
different kinds of fields including ``fast quantum information
processing'', ``fast cold-atom'', ``fast ion transport'', ``fast
wave-packet splitting'', ``fast expansion'', etc.
\cite{c1,ap7,ap8,ap9,ap10,ap11,ap12,ap13,ap14,ap15,ap16,ap17,ap18,ap19,ap20,ap21,cc1,ymd1,ymd2,ymd3}.
For example, with invariant-based inverse engineering, a fast
population transfer in a three-level system has been achieved by
Chen and Muga \cite{c1}. Chen \emph{et al.} \cite{cc1} have proposed
a scheme for fast generation of three-atom singlet states by TQD.
These schemes 
have shown the
powerful application for invariant-based inverse engineering and TQD
in QIP.

It has been pointed out in Ref. \cite{sa4} that, invariant-based
inverse engineering and TQD are strongly related and potentially
equivalent to each other. Invariant-based method is convenience and
effective with a Hamiltonian which admits known structures for the
invariants. But for most systems, the invariants are unknown or hard
to be solved. As for TQD, it will not meet this difficult point.
However, some terms of Hamiltonian constructed by TQD, which are
difficult to be realized in experiments, may appear when we
accelerate adiabatic schemes. Therefore, how to avoid these
problematic terms is a notable problem. Till now, some schemes
\cite{add2,add3,add4,add5,add1,add10,add6,add7,add8,add9} have been
proposed to solve the problem of the TQD method recently. For
example, Ib\'{a}\~{n}ez \emph{et al.} \cite{add5} have produced a
sequence of STA by examining the limitations and capabilities of
superadiabatic iterations. Ib\'{a}\~{n}ez \emph{et al.} \cite{add1}
have also studied the STA for a two-level system with multiple
Schr\"{o}dinger pictures, and subsequently, Song \emph{et al.}
\cite{add10} have expanded the method in a three-level system based
on two nitrogen-vacancy-center ensembles coupled to a transmission
line resonator. Moreover, without directly using the counterdiabatic
Hamiltonian, Torrontegui \emph{et al.} \cite{add6} have used the
dynamical symmetry of the Hamiltonian to find alternative
Hamiltonians that achieved the same goals as speed-up schemes with
Lie transforms. Chen \emph{et al.} \cite{add9} have proposed a
method for constructing shortcuts to adiabaticity by a substitute of
counterdiabatic driving terms.

In this paper, inspired by TQD and the previous schemes
\cite{add1,add2,add3,add4,add5,add6,add7,add8,add9,add10}, a new
scheme for reverse engineering of a Hamiltonian by designing the
evolution operators is proposed for eliminating the terms of
Hamiltonian which are hard to be realized in practice. The present
scheme is focus on only one or parts of moving states in a
$D$-dimension ($D\geq3$) system, that is different from TQD with
which all instantaneous eigenstates evolve parallel. According to
the numerical simulation, the present scheme not only contains the
results of TQD, but also has more free parameters, which make this
scheme more flexible. Moreover, the problematic terms of Hamiltonian
may be eliminated by suitably choosing these new free parameters.
For the sake of clearness, an example is given to realize the
population transfer for a Rydberg atom, where numerical simulation
shows the scheme is effective. Therefore, this scheme may be used to
construct a Hamiltonian which can be realized in experiments.

The article is organized as follows. In the section of ``Reverse
engineering of a Hamiltonian'', we will introduce the basic
principle of the scheme for reverse engineering of a Hamiltonian by
designing the evolution operators. In the section of ``The
population transfer for a Rydberg atom'', we will show an example
using the present scheme to realize the population transfer for a
Rydberg atom. Finally, conclusions will be given in the section of
``Conclusion''.

\section*{Reverse engineering of a Hamiltonian}

We begin to introduce the basic method of the scheme for reverse
engineering of a Hamiltonian by designing the evolution operators.
Firstly, we suppose that the system evolves along the state
$|\phi_1(t)\rangle$ and the initial state of the system is
$|\psi(0)\rangle$. So, the condition
$|\phi_1(0)\rangle=|\psi(0)\rangle$ should be satisfied. We can
obtain a complete orthogonal basis $\{|\phi_n(t)\rangle\}$ through a
process of completion and orthogonalization. Therefore, the vectors
in basis $\{|\phi_n(t)\rangle\}$ satisfy the orthogonality condition
$\langle\phi_m(t)|\phi_n(t)\rangle=\delta_{mn}$ and the completeness
condition $\sum\limits_{n}|\phi_n(t)\rangle\langle\phi_{n}(t)|=1$.
Since the system evolves along $|\phi_1(t)\rangle$, the evolution
operator can be designed as
\begin{eqnarray}\label{bm1}
U(t)=|\phi_1(t)\rangle\langle\phi_1(0)|+\sum\limits_{m,n\neq1}\lambda_{mn}(t)|\phi_m(t)\rangle\langle\phi_n(0)|,
\end{eqnarray}
where parameters $\lambda_{mn}(t)$ ($m,n\neq1$) are chosen to
satisfy the unitary condition $UU^{\dag}=U^{\dag}U=1$. Submitting
the unitary condition into Eq.~(\ref{bm1}), we obtain
\begin{eqnarray}\label{bm2}
\sum\limits_{k\neq1}\lambda_{mk}(t)\lambda^*_{nk}(t)=\delta_{mn}\
(m,n\neq1).
\end{eqnarray}

Secondly, according to Schr\"{o}dinger equation ($\hbar=1$), we have
\begin{eqnarray}\label{bm3}
&&i\partial_t|\psi(t)\rangle=H(t)|\psi(t)\rangle,\cr\cr&&
i\partial_tU(t)|\psi(0)\rangle=H(t)U(t)|\psi(0)\rangle.
\end{eqnarray}
On account of the arbitrariness of $|\psi(0)\rangle$,
Eq.~(\ref{bm3}) can be written by
\begin{eqnarray}\label{bm4}
i\partial_tU(t)=H(t)U(t).
\end{eqnarray}
The Hamiltonian can be formally solved from Eq.~(\ref{bm4}), and be
given as
\begin{eqnarray}\label{bm5}
H(t)&=&i(\partial_tU(t))U^{\dag}(t)\cr\cr&&
=i|\dot{\phi}_1(t)\rangle\langle\phi_1(t)|+i\sum\limits_{l,m,n\neq1}\lambda_{ml}(t)\lambda^*_{nl}(t)|\dot{\phi}_m(t)\rangle\langle\phi_n(t)|\cr\cr&&
+i\sum\limits_{l,m,n\neq1}\dot{\lambda}_{ml}(t)\lambda^*_{nl}(t)|\phi_m(t)\rangle\langle\phi_n(t)|.
\end{eqnarray}
By submitting Eq.~(\ref{bm2}) into Eq.~(\ref{bm5}), the Hamiltonian
in Eq.~(\ref{bm5}) can be described as
\begin{eqnarray}\label{bm6}
H(t)=i\sum\limits_{k}|\dot{\phi}_k(t)\rangle\langle\phi_k(t)|+i\sum\limits_{l,m,n\neq1}\dot{\lambda}_{ml}(t)\lambda^*_{nl}(t)|\phi_m(t)\rangle\langle\phi_n(t)|.
\end{eqnarray}
Different from TQD, which gives Hamiltonian in the following from
\begin{eqnarray}\label{bm7}
H(t)=i\sum\limits_{k}|\dot{\phi}_k(t)\rangle\langle\phi_k(t)|,
\end{eqnarray}
the present scheme has more free parameters $\lambda_{mn}(t)$.
Therefore, this scheme may construct some new and different
Hamiltonians. 
Moreover, when parameters $\lambda_{mn}$ ($m,n\neq1$) are
independent of time, Eq.~(\ref{bm6}) will degenerate into
Eq.~(\ref{bm7}), which shows that the present scheme contains the
results of TQD. On the other hand, once the unitary condition
$UU^{\dag}=U^{\dag}U=1$ for evolution operator is satisfied, the
Hamiltonian given in Eq.~(\ref{bm6}) should be a Hermitian operator,
because
\begin{eqnarray}\label{bm8}
H(t)&=&i(\partial_tU(t))U^{\dag}(t)\cr\cr&&
=i\partial_t(U(t)U^{\dag}(t))-i U(t)\partial_t(U^{\dag}(t))\cr\cr&&
=-i U(t)\partial_t(U^{\dag}(t))\cr\cr&& =H^{\dag}(t).
\end{eqnarray}

As an extension, for a $N$-dimension system ($N\geq4$), the
evolution operator can be designed as
\begin{eqnarray}\label{bm9}
U(t)=\sum\limits_{j=1}^{s}|\phi_j(t)\rangle\langle\phi_j(0)|+\sum\limits_{\mathop{m,n\neq
j}\limits_{j=1,2,\cdots,s}}\lambda_{mn}(t)|\phi_m(t)\rangle\langle\phi_n(0)|,\
(1\leq s\leq N-2) .
\end{eqnarray}
Then, the initial state $|\psi(0)\rangle$ of the system can be
expressed by the superposition of $\{|\phi_j(0)\rangle\}$
$(j=1,2,\cdots,s)$. Thus, the system can evolve along more than one
moving states in this case. This might sometimes help us to simplify
the design of the system's Hamiltonian.

\section*{The
population transfer for a Rydberg atom}

For the sake of clearness, we give an example to emphasize the
advantages of the scheme. Here, we consider a Rydberg atom with the
energy levels shown in Fig. 1. The transition between $|1\rangle$
and $|3\rangle$ is hard to realize. So, the Hamiltonian of the
Rydberg atom is usually written as the following form
\begin{equation}\label{e0}
H(t)=\Omega_{12}(t)|1\rangle\langle2|
+\Omega_{23}(t)e^{i\varphi(t)}|2\rangle\langle3|+H.c.,
\end{equation}
where, $\Omega_{12}$ and $\Omega_{23}$ are the Rabi frequencies of
laser pulses, which drive the transitions
$|1\rangle\leftrightarrow|2\rangle$ and
$|2\rangle\leftrightarrow|3\rangle$, respectively, and they are
$\varphi$-dephased from each other. Suppose the initial state of the
three-energy-level Rydberg atom is $|1\rangle$, the target state is
$|\Psi_{tar}\rangle=\cos\mu|1\rangle+\sin\mu|3\rangle$. We choose a
complete orthogonal basis as below

\begin{eqnarray}\label{e1}
&&|\phi_1(t)\rangle=\cos\alpha\cos\beta|1\rangle+\sin\beta|2\rangle+\sin\alpha\cos\beta|3\rangle,\cr\cr&&
|\phi_2(t)\rangle=\cos\alpha\sin\beta|1\rangle-\cos\beta|2\rangle+\sin\alpha\sin\beta|3\rangle,\cr\cr&&
|\phi_3(t)\rangle=\sin\alpha|1\rangle-\cos\alpha|3\rangle.
\end{eqnarray}

  With the unitary condition in Eq.~(\ref{bm2}), the evolution
operator can take this form
\begin{eqnarray}\label{e2}
U(t)&=&|\phi_1(t)\rangle\langle\phi_1(0)|+\cos\lambda(t)(|\phi_2(t)\rangle\langle\phi_2(0)|+|\phi_3(t)\rangle\langle\phi_3(0)|)\cr\cr&&
+\sin\lambda(t)(e^{i\theta(t)}|\phi_2(t)\rangle\langle\phi_3(0)|-e^{-i\theta(t)}|\phi_3(t)\rangle\langle\phi_2(0)|).
\end{eqnarray}
According to Eq.~(\ref{bm6}), the evolution operator in
Eq.~(\ref{e2}) gives the following Hamiltonian
\begin{eqnarray}\label{e3}
H(t)&=&i\sum\limits_{k=1}^{3}|\dot{\phi}_k(t)\rangle\langle\phi_k(t)|+i\dot\lambda(e^{i\theta}|\phi_2(t)\rangle\langle\phi_3(t)|-e^{i\theta}|\phi_3(t)\rangle\langle\phi_2(t)|)\cr\cr&&
-\dot{\theta}\sin\lambda\cos\lambda(e^{i\theta}|\phi_2(t)\rangle\langle\phi_3(t)|+e^{-i\theta}|\phi_3(t)\rangle\langle\phi_2(t)|)\cr\cr&&
-\dot{\theta}\sin^2\lambda(|\phi_2(t)\rangle\langle\phi_2(t)|-|\phi_3(t)\rangle\langle\phi_3(t)|).
\end{eqnarray}
For simplicity, we set $\theta=0$ here, the Hamiltonian in
Eq.~(\ref{e3}) can be written by
\begin{eqnarray}\label{e4}
H(t)&=&i(\dot{\lambda}\sin\beta+\dot{\alpha})(|3\rangle\langle1|-|1\rangle\langle3|)\cr\cr&&
+i(\dot{\beta}\cos\alpha-\dot{\lambda}\cos\beta\sin\alpha)(|2\rangle\langle1|-|1\rangle\langle2|)\cr\cr&&
+i(\dot{\beta}\sin\alpha+\dot{\lambda}\cos\alpha\cos\beta)(|2\rangle\langle3|-|3\rangle\langle2|).
\end{eqnarray}
Here, the Hamiltonian in Eq.~(\ref{e4}) is already a Hermitian
operator. To eliminate the terms with $|1\rangle\langle3|$ and
$|3\rangle\langle1|$, which are difficult to realize for the
three-energy-level Rydberg atom, we set
$\dot{\lambda}\sin\beta+\dot{\alpha}=0$. Eq.~(\ref{e4}) will be
changed into
\begin{eqnarray}\label{e5}
H(t)&=&i\Omega_1(t)(|2\rangle\langle1|-|1\rangle\langle2|)\cr\cr&&
+i\Omega_2(t)(|2\rangle\langle3|-|3\rangle\langle2|),\cr\cr&&
\Omega_1(t)=\dot{\beta}\cos\alpha+\dot{\alpha}\cot\beta\sin\alpha,\cr\cr&&
\Omega_2(t)=\dot{\beta}\sin\alpha-\dot{\alpha}\cos\alpha\cot\beta.
\end{eqnarray}

For simplicity, we suppose the initial time is $t_i=0$ and the final
time is $t_f=T$, so $T$ is the total interaction time. To satisfy
the boundary conditions $\alpha(0)=0$, $\alpha(T)=\mu$,
$\dot{\alpha}(0)=\dot{\alpha}(T)=0$, $\beta(0)=\beta(T)=0$,
$\dot{\beta}(0)=\dot{\beta}(T)=0$ and avoid the singularity of
Hamiltonian, we choose the parameters as
\begin{eqnarray}\label{e6}
&&\beta(t)=\frac{A}{2}[1-\cos(\frac{2\pi t}{T})],\cr\cr&&
\dot{\beta}(t)=\frac{\pi A}{T}\sin(\frac{2\pi t}{T}),\cr\cr&&
\dot{\alpha}(t)=\frac{8\mu}{3T}\sin^4(\frac{\pi t}{T}),\cr\cr&&
\alpha(t)=\mu\frac{t}{T}-\frac{2\mu}{3\pi}\sin(\frac{2\pi
t}{T})+\frac{\mu}{12\pi}\sin(\frac{4\pi t}{T}),
\end{eqnarray}
where $A$ is an arbitrary constant. Then, the Hamiltonian in
Eq.~(\ref{e5}) can be written by
\begin{eqnarray}\label{e7}
H(t)&=&i\Omega_1(t)(|2\rangle\langle1|-|1\rangle\langle2|)
+i\Omega_2(t)(|2\rangle\langle3|-|3\rangle\langle2|),\cr\cr&&
\Omega_1(t)=\frac{\pi A}{T}\sin(\frac{2\pi
t}{T})\cos\alpha+\frac{2\mu}{3T}[1-\cos(\frac{2\pi
t}{T})]^2\sin\alpha\cot\beta,\cr\cr&& \Omega_2(t)=\frac{\pi
A}{T}\sin(\frac{2\pi
t}{T})\sin\alpha-\frac{2\mu}{3T}[1-\cos(\frac{2\pi
t}{T})]^2\cos\alpha\cot\beta.
\end{eqnarray}

For the sake of obtaining a relatively high speed, the values of
$\Omega_1T$ and $\Omega_2T$ in Eq.~(\ref{e7}) should not be too
large. Noticing that, with $A$ increasing, $\pi A$ increases while
$\cot\beta$ decreases. Therefore, to obtain a relatively small
$|\Omega_1T|$ and $|\Omega_2T|$, $A$ should be neither too large nor
too small. Therefore, we choose $A=1$ here. However, we can see from
Eq.~(\ref{e7}) that the functions of Rabi frequencies $\Omega_1(t)$
and $\Omega_2(t)$ are too complex for experimental realization.
Fortunately, we can solve the problem by using simple functions to
make a curve fitting for the $\Omega_1(t)$ and $\Omega_2(t)$. As an
example, $\mu=\frac{\pi}{4}$ is taken here. We use $\Omega'_1(t)$
and $\Omega'_2(t)$ in the following, which are linear superposition
of the Gaussian or trigonometric functions, to make a curve fitting
for the $\Omega_1(t)$ and $\Omega_2(t)$,
\begin{eqnarray}\label{e8}
H(t)&=&i\Omega_1'(t)(|2\rangle\langle1|-|1\rangle\langle2|)
+i\Omega_2'(t)(|2\rangle\langle3|-|3\rangle\langle2|),\cr\cr&&
\Omega_1'(t)=\begin{cases} \frac{3.154}{T}\sin(5.939t/T-0.02523),
&0\leq t\leq 0.534T, \cr \frac{1.686}{T}\sin(6.531t/T-0.3177),
&0.534T\leq t\leq T,
\end{cases}\cr\cr&&
\Omega_2'(t)=-\frac{1}{T}[0.9443e^{-(\frac{t-0.3185T}{0.1848T})^2}+2.95e^{-(\frac{t-0.7233T}{0.2004T})^2}].
\end{eqnarray}
In this case, we have $|\Omega_1'T|\leq3.154$ and
$|\Omega_2'T|\leq2.96$.

To compare the values of $\Omega_1(t)$ and $\Omega_1'(t)$,
$\Omega_2(t)$ and $\Omega_2'(t)$, we plot $\Omega_1T$ and
$\Omega_1'T$ versus $t/T$ with $\mu=\pi/4$ and $A=1$ in Fig. 2 (a)
and plot $\Omega_2T$ and $\Omega_2'T$ versus $t/T$ with $\mu=\pi/4$
and $A=1$ in Fig. 2 (b). From Figs. 2 (a) and (b), one can find that
the curves of $\Omega_1(t)$ and $\Omega_1'(t)$ ($\Omega_2(t)$ and
$\Omega_2'(t)$) are well matched with each other. Therefore, we may
use $\Omega'_1(t)$ ($\Omega'_2(t)$) instead of $\Omega_1(t)$
($\Omega_2(t)$) to obtain the same effect. To test the effectiveness
of the approximation by using $\Omega'_1(t)$ ($\Omega'_2(t)$)
instead of $\Omega_1(t)$ ($\Omega_2(t)$), a simulation for the
varies of populations of states $|1\rangle$, $|2\rangle$ and
$|3\rangle$ when the Rydberg atom is driven by laser pulses with
Rabi frequencies $\Omega_1(t)$ and $\Omega_2(t)$ with parameters
$\mu=\pi/4$ and $A=1$, is shown in Fig. 3 (a). We can see from Fig.
3 (a) that the evolution is consonant with the expectation coming
from the evolution operator in Eq.~(\ref{e2}). As a comparison, a
simulation for the varies of populations of states $|1\rangle$,
$|2\rangle$ and $|3\rangle$ when the Rydberg atom is driven by laser
pulses with Rabi frequencies $\Omega'_1(t)$ and $\Omega'_2(t)$ with
parameters $\mu=\pi/4$ and $A=1$, is shown in Fig. 3 (b). As shown
in Fig. 3 (a) and Fig. 3 (b), we can conclude that the approximation
by using $\Omega'_1(t)$ ($\Omega'_2(t)$) instead of $\Omega_1(t)$
($\Omega_2(t)$) is effective here. In addition, seen from Fig. 3,
the population of intermediate state $|2\rangle$ reaches a peak
value about 0.72, because the system does not evolve along the dark
state of the Hamiltonian of the system but a nonadiabatic shortcut,
which greatly reduces the total evolution time.

  Since most of the parameters are hard to faultlessly achieve in
experiment, that require us to investigate the variations in the
parameters caused by the experimental imperfection. We would like to
discuss the fidelity $F=|\langle\Psi_{tar}|\phi_1(T)\rangle|^2$ with
the deviations $\delta T$, $\delta\Omega_1'$ and $\delta\Omega_2'$
of total interaction time $T$, Rabi frequencies of laser pulses
$\Omega_1'$ and $\Omega_2'$ being considered.

  Firstly, we plot $F$ versus $\delta\Omega_1'/\Omega_1'$ and
$\delta\Omega_2'/\Omega_2'$ with parameters $\mu=\pi/4$ and $A=1$ in
Fig. 4 (a). Moreover, we calculate the exact values of the
fidelities $F$ at some boundary points of Fig. 4 (a) and show the
results in Table I. According to Table I and Fig. 4 (a), we find
that the final fidelity $F$ is still higher than 0.9822 even when
the deviation
$|\delta\Omega_1'/\Omega_1'|=|\delta\Omega_2'/\Omega_2'|=10\%$.
Therefore, the realizing of the population transfer for a Rydberg
atom given in this paper is robust against deviations
$\delta\Omega_1'$ and $\delta\Omega_2'$ of Rabi frequencies
$\Omega_1'$ and $\Omega_2'$ for laser pulses.

  Secondly, we plot $F$ versus $\delta\Omega_1'/\Omega_1'$ and $\delta
T/T$ with parameters $\mu=\pi/4$ and $A=1$ in Fig. 4 (b). Moreover,
$\delta\Omega_1'/\Omega_1'$ and $\delta T/T$ with corresponding
fidelity $F$ are shown in Table II. Seen from Table II and Fig. 4
(b), we obtain that the fidelity $F$ is still high than 0.9729 even
when the deviation $|\delta\Omega_1'/\Omega_1'|=|\delta T/T|=10\%$.
So, the scheme is insensitive to deviations $\delta\Omega_1'$ and
$\delta T$.

  Thirdly, $F$ versus $\delta\Omega_2'/\Omega_2'$ and $\delta
T/T$  with parameters $\mu=\pi/4$ and $A=1$ is plotted in Fig. 4
(c). And $\delta\Omega_2'/\Omega_2'$ and $\delta T/T$ with
corresponding fidelity $F$ are given in Table III. As indicated in
Table III and Fig. 4 (c), the fidelity $F$ is still high than 0.9588
even when the deviation $|\delta\Omega_2'/\Omega_2'|=|\delta
T/T|=10\%$. Moreover, when deviations of $\delta\Omega_2'$ and
$\delta T$ have the different signs (one negative and one positive),
the fidelity $F$ can still keep in a high level. Hence, we can say
the scheme suffers little from deviations $\delta\Omega_2'$ and
$\delta T$.

  Fourthly, we discuss the fidelity $F$ when $\delta\Omega_1'$,
$\delta\Omega_2'$ and $\delta T$ are all considered. Some samples
are given in Table IV. Table IV shows that the fidelity $F$ is still
with a high level when the three deviations $\delta\Omega_1'$,
$\delta\Omega_2'$ and $\delta T$ are all considered. Moreover, in
the worst case, when
$\delta\Omega_1'/\Omega_1'=\delta\Omega_2'/\Omega_2'=\delta
T/T=-10\%$, the fidelity $F$ is still higher than 0.9469.

  According to the analysis above, we summarize that, the scheme to
realize the population transfer for a Rydberg atom is robust against
operational imperfection.

  To prove that the present scheme can be used to speed up the
system's evolution and construct the shortcut to adiabatic passages,
we make a comparison between the present scheme and the fractional
stimulated Raman adiabatic passage (STIRAP) method via dark state
$|\Psi_{dark}(t)\rangle=\frac{1}{\sqrt{\Omega_{12}^2(t)+\Omega_{23}^2(t)}}(\Omega_{23}(t)|1\rangle-\Omega_{12}(t)|3\rangle)$
of Hamiltonian shown in Eq.~(\ref{e0}). According to STIRAP method,
by setting boundary condition
\begin{eqnarray}\label{e9}
\lim\limits_{t\rightarrow-\infty}=\frac{\Omega_{12}(t)}{\Omega_{23}(t)}=0,\
\lim\limits_{t\rightarrow+\infty}=\frac{\Omega_{12}(t)}{\Omega_{23}(t)}=-\tan\mu=-1,
\end{eqnarray}
one can design the Rabi frequencies $\Omega_{12}(t)$ and
$\Omega_{23}(t)$ as following
\begin{eqnarray}\label{e10}
&&\Omega_{12}(t)=-\Omega_0\exp[-(\frac{t-t_0-T/2}{t_c})^2]\sin\mu,\cr\cr&&
\Omega_{23}(t)=\Omega_0\exp[-(\frac{t+t_0-T/2}{t_c})^2]+\Omega_0\exp[-(\frac{t-t_0-T/2}{t_c})^2]\cos\mu,
\end{eqnarray}
where $\Omega_0$ denotes the pulse amplitude, $t_c$ and $t_0$ are
some related parameters. Setting $t_c=0.19t_f$ and $t_0=0.14t_f$,
Rabi frequencies $\Omega_{12}(t)$ and $\Omega_{23}(t)$ can well
satisfy the boundary condition in Eq.~(\ref{e9}). We plot Fig. 5 to
show the fidelity $F$ when the Rydberg atom is driven by laser
pulses with Rabi frequencies $\Omega_{12}(t)$ and $\Omega_{23}(t)$
shown in Eq.~(\ref{e10}) versus $\Omega_0T$. And a series of samples
of $\Omega_0T$ and corresponding fidelity $F$ are shown in Table V.
From Fig. 5 and Table V, we can see that, to meet the adiabatic
condition and obtain a relatively high fidelity by using STIRAP
method, one should take $\Omega_0T$ about 30. Moreover, when
$\Omega_0T=3.154$, the adiabatic condition is badly violated and the
fidelity is only 0.5538 for STIRAP method. But for the present
scheme, we can obtain $F=1.000$ while $|\Omega_1'T|\leq3.154$ and
$|\Omega_2'T|\leq2.96$. Therefore, the evolution speed with the
present scheme is faster a lot comparing with that using STIRAP
method. It confirms that the present scheme can be used to speed up
the system's evolution and construct the shortcut to adiabatic
passages. Therefore, we conclude that the present scheme can
construct a Hamiltonian with both fast evolution process and
robustness against operational imperfection.

  In the end, we discuss the fidelity $F$ is robust to the decoherence mechanisms.
In this scheme, the atomic spontaneous emission plays the major
role. The evolution of the system can be described by a master
equation in Lindblad form as following
\begin{equation}\label{d1}
\dot{\rho}=i[\rho,H_I]+\sum\limits_{l}[L_l\rho
L_l^{\dagger}-\frac{1}{2}(L_l^{\dagger}L_l\rho+\rho
L_l^{\dagger}L_l)],
\end{equation}
where, $L_l$ is the Lindblad operator. There are two Lindblad
operators here. They are $L_{1}=\sqrt{\Gamma_1}|1\rangle\langle 2|$
and $L_{2}=\sqrt{\Gamma_2}|2\rangle\langle 3|$, in which, $\Gamma_1$
and $\Gamma_2$ are the atomic spontaneous emission coefficients for
$|2\rangle\rightarrow|1\rangle$ and $|3\rangle\rightarrow|2\rangle$,
respectively. Fidelity $F$ versus $\Gamma_1T$ and $\Gamma_2T$ is
plotted in Fig. 6. From Fig. 6, we can see that the fidelity $F$
decreases when $\Gamma_1$ and $\Gamma_2$ increase. When in the case
of strong coupling $\Omega_1,\Omega_2\gg\Gamma_1,\Gamma_2$, the
influence caused by atomic spontaneous emission is little. For
example, if $\Gamma_1=\Gamma_2=0.01\times3.154/T$, the fidelity is
0.9901. Even when $\Gamma_1=\Gamma_2=0.1\times3.154/T$, the fidelity
is 0.9101, still higher than 0.9. With current experimental
technology, it is easy to obtain a laser pulse with Rabi frequency
much larger than the atomic spontaneous emission coefficients.
Therefore, the population transfer for a Rydberg atom with the
reverse engineering scheme given here can be robustly realized.

\section*{Conclusion}

In conclusion, we have proposed an effective and flexible scheme for
reverse engineering of a Hamiltonian by designing the evolution
operators. Different from TQD, the present scheme is focus on only
one or parts of moving states in a $D$-dimension ($D\geq3$) system.
The numerical simulation has indicated that the present scheme not
only contains the results of TQD, but also has more free parameters,
which make this scheme more flexible. Moreover, the new free
parameters may help to eliminate the terms of Hamiltonian which are
hard to be realized practically. Furthermore, owing to suitable
choice of boundary conditions for parameters, by making a curve
fitting, the complex Rabi frequencies $\Omega_1$ and $\Omega_2$ of
laser pulses can be respective superseded by Rabi frequencies
$\Omega_1'$ and $\Omega_2'$ expressed by the superpositions of the
Gaussian or trigonometric functions, which can be realized with
current experimental technology. The example given in Sec. III has
shown that the present scheme can design a Hamiltonian to realize
the population transfer for a Rydberg atom successfully and the
numerical simulation has shown that the scheme is fast and
robustness against the operational imperfection and the decoherence
mechanisms. Therefore, the present scheme may be used to construct a
Hamiltonian which can be realized in experiments.

\section*{Acknowledgement}

This work was supported by the National Natural Science Foundation
of China under Grants No. 11575045 and No. 11374054, and the Major
State Basic Research Development Program of China under Grant No.
2012CB921601.

\section*{Author Contributions}

Y. X. and Y. H. K. came up with the initial idea for the work and
performed the simulations for the model. Y. H. C., Q. C. W., J. S.
and B. H. H. performed the calculations for the model.  Y. X., Y. H.
K. and Y. H. C. performed all the data analysis and the initial
draft of the manuscript. All authors participated in the writing and
revising of the text.

\section*{Additional Information}

 Competing financial interests: The authors
declare no competing financial interests.

\newpage

\begin{center}{\bf Table I. $\delta\Omega_1'/\Omega_1'$ and $\delta\Omega_2'/\Omega_2'$ with corresponding fidelity $F$.\ \ \ \ \ \ \ \ \ \ \ \
\ \ \ \ \ \ \ \ \ \ \  }{\small
\begin{tabular}{ccc}\hline\hline
$\delta\Omega_1'/\Omega_1'$ \ \ \ \ \ \ \ \ \ \ \ \ \ \ \ \ \ \ \ \ \ \ \ &$\delta\Omega_2'/\Omega_2'$ \ \ \ \ \ \ \ \ \ \ \ \ \ \ \ \ \ \ \ \ \ &$F$\\
\hline $10\%\ \ \ \ \ \ \ \ \ \ \ \ \ \ \ \ \ \ \ \ \ \ \ \ $&$
10\%\
\ \ \ \ \ \ \ \ \ \ \ \ \ \ \ $&$0.9835$\\
$10\%\ \ \ \ \ \ \ \ \ \ \ \ \ \ \ \ \ \ \ \ \ \ \ \ $&$ 0\ \ \ \ \
\
\ \ \ \ \ \ \ \ \ \ $&$0.9951$\\
$0\ \ \ \ \ \ \ \ \ \ \ \ \ \ \ \ \ \ \ \ \ \ \ \ $&$ 10\%\ \ \ \ \
\
\ \ \ \ \ \ \ \ \ \ $&$0.9916$\\
$0\ \ \ \ \ \ \ \ \ \ \ \ \ \ \ \ \ \ \ \ \ \ \ \ $&$ 0\ \ \ \ \ \ \
\ \ \ \ \ \ \ \ \ $&$1.0000$\\
$-10\%\ \ \ \ \ \ \ \ \ \ \ \ \ \ \ \ \ \ \ \ \ \ \ \ $&$ 0\ \ \ \ \
\ \ \ \ \ \ \ \ \ \ \ $&$0.9938$\\
$0\ \ \ \ \ \ \ \ \ \ \ \ \ \ \ \ \ \ \ \ \ \ \ \ $&$ -10\%\ \ \ \ \
\ \ \ \ \ \ \ \ \ \ \ $&$0.9902$\\
$-10\%\ \ \ \ \ \ \ \ \ \ \ \ \ \ \ \ \ \ \ \ \ \ \ \ $&$ -10\%\ \ \
\
\ \ \ \ \ \ \ \ \ \ \ \ $&$0.9822$\\
$10\%\ \ \ \ \ \ \ \ \ \ \ \ \ \ \ \ \ \ \ \ \ \ \ \ $&$ -10\%\ \ \
\
\ \ \ \ \ \ \ \ \ \ \ \ $&$0.9875$\\
$-10\%\ \ \ \ \ \ \ \ \ \ \ \ \ \ \ \ \ \ \ \ \ \ \ \ $&$ 10\%\ \ \
\
\ \ \ \ \ \ \ \ \ \ \ \ $&$0.9887$\\
\hline \hline
\end{tabular} }
\end{center}

\begin{center}{\bf Table II. $\delta\Omega_1'/\Omega_1'$ and $\delta T/T$ with corresponding fidelity $F$.\ \ \ \ \ \ \ \ \ \ \ \
\ \ \ \ \ \ \ \ \ \ \  }{\small
\begin{tabular}{ccc}\hline\hline
$\delta\Omega_1'/\Omega_1'$ \ \ \ \ \ \ \ \ \ \ \ \ \ \ \ \ \ \ \ \ \ \ \ &$\delta T/T$ \ \ \ \ \ \ \ \ \ \ \ \ \ \ \ \ \ \ \ \ \ &$F$\\
\hline $10\%\ \ \ \ \ \ \ \ \ \ \ \ \ \ \ \ \ \ \ \ \ \ \ \ $&$
10\%\
\ \ \ \ \ \ \ \ \ \ \ \ \ \ \ $&$0.9855$\\
$10\%\ \ \ \ \ \ \ \ \ \ \ \ \ \ \ \ \ \ \ \ \ \ \ \ $&$ 0\ \ \ \ \
\
\ \ \ \ \ \ \ \ \ \ $&$0.9951$\\
$0\ \ \ \ \ \ \ \ \ \ \ \ \ \ \ \ \ \ \ \ \ \ \ \ $&$ 10\%\ \ \ \ \
\
\ \ \ \ \ \ \ \ \ \ $&$0.9942$\\
$0\ \ \ \ \ \ \ \ \ \ \ \ \ \ \ \ \ \ \ \ \ \ \ \ $&$ 0\ \ \ \ \ \ \
\ \ \ \ \ \ \ \ \ $&$1.0000$\\
$-10\%\ \ \ \ \ \ \ \ \ \ \ \ \ \ \ \ \ \ \ \ \ \ \ \ $&$ 0\ \ \ \ \
\ \ \ \ \ \ \ \ \ \ \ $&$0.9938$\\
$0\ \ \ \ \ \ \ \ \ \ \ \ \ \ \ \ \ \ \ \ \ \ \ \ $&$ -10\%\ \ \ \ \
\ \ \ \ \ \ \ \ \ \ \ $&$0.9855$\\
$-10\%\ \ \ \ \ \ \ \ \ \ \ \ \ \ \ \ \ \ \ \ \ \ \ \ $&$ -10\%\ \ \
\
\ \ \ \ \ \ \ \ \ \ \ \ $&$0.9729$\\
$10\%\ \ \ \ \ \ \ \ \ \ \ \ \ \ \ \ \ \ \ \ \ \ \ \ $&$ -10\%\ \ \
\
\ \ \ \ \ \ \ \ \ \ \ \ $&$0.9879$\\
$-10\%\ \ \ \ \ \ \ \ \ \ \ \ \ \ \ \ \ \ \ \ \ \ \ \ $&$ 10\%\ \ \
\
\ \ \ \ \ \ \ \ \ \ \ \ $&$0.9915$\\
\hline \hline
\end{tabular} }
\end{center}

\begin{center}{\bf Table III. $\delta\Omega_2'/\Omega_2'$ and $\delta T/T$ with corresponding fidelity $F$.\ \ \ \ \ \ \ \ \ \ \ \
\ \ \ \ \ \ \ \ \ \ \  }{\small
\begin{tabular}{ccc}\hline\hline
$\delta\Omega_2'/\Omega_2'$ \ \ \ \ \ \ \ \ \ \ \ \ \ \ \ \ \ \ \ \ \ \ \ &$\delta T/T$ \ \ \ \ \ \ \ \ \ \ \ \ \ \ \ \ \ \ \ \ \ &$F$\\
\hline $10\%\ \ \ \ \ \ \ \ \ \ \ \ \ \ \ \ \ \ \ \ \ \ \ \ $&$
10\%\
\ \ \ \ \ \ \ \ \ \ \ \ \ \ \ $&$0.9688$\\
$10\%\ \ \ \ \ \ \ \ \ \ \ \ \ \ \ \ \ \ \ \ \ \ \ \ $&$ 0\ \ \ \ \
\
\ \ \ \ \ \ \ \ \ \ $&$0.9916$\\
$0\ \ \ \ \ \ \ \ \ \ \ \ \ \ \ \ \ \ \ \ \ \ \ \ $&$ 10\%\ \ \ \ \
\
\ \ \ \ \ \ \ \ \ \ $&$0.9942$\\
$0\ \ \ \ \ \ \ \ \ \ \ \ \ \ \ \ \ \ \ \ \ \ \ \ $&$ 0\ \ \ \ \ \ \
\ \ \ \ \ \ \ \ \ $&$1.0000$\\
$-10\%\ \ \ \ \ \ \ \ \ \ \ \ \ \ \ \ \ \ \ \ \ \ \ \ $&$ 0\ \ \ \ \
\ \ \ \ \ \ \ \ \ \ \ $&$0.9902$\\
$0\ \ \ \ \ \ \ \ \ \ \ \ \ \ \ \ \ \ \ \ \ \ \ \ $&$ -10\%\ \ \ \ \
\ \ \ \ \ \ \ \ \ \ \ $&$0.9855$\\
$-10\%\ \ \ \ \ \ \ \ \ \ \ \ \ \ \ \ \ \ \ \ \ \ \ \ $&$ -10\%\ \ \
\
\ \ \ \ \ \ \ \ \ \ \ \ $&$0.9588$\\
$10\%\ \ \ \ \ \ \ \ \ \ \ \ \ \ \ \ \ \ \ \ \ \ \ \ $&$ -10\%\ \ \
\
\ \ \ \ \ \ \ \ \ \ \ \ $&$0.9974$\\
$-10\%\ \ \ \ \ \ \ \ \ \ \ \ \ \ \ \ \ \ \ \ \ \ \ \ $&$ 10\%\ \ \
\
\ \ \ \ \ \ \ \ \ \ \ \ $&$0.9994$\\
\hline \hline
\end{tabular} }
\end{center}

\begin{center}{\bf Table IV. $\delta\Omega_1'/\Omega_1'$, $\delta\Omega_2'/\Omega_2'$ and $\delta T/T$ with corresponding fidelity $F$.\ \ \ \ \ \ \ \ \ \ \ \
\ \ \ \ \ \ \ \ \ \ \  }{\small
\begin{tabular}{cccc}\hline\hline
$\delta\Omega_1'/\Omega_1'$ \ \ \ \ \ \ \ \ \ \ \ \ \ \ \ \ \ \ \ \ \ \ \ &$\delta\Omega_2'/\Omega_2'$ \ \ \ \ \ \ \ \ \ \ \ \ \ \ \ \ \ \ \ \ \ &$\delta T/T$ \ \ \ \ \ \ \ \ \ \ \ \ \ \ \ \ \ \ \ \ \ \ \ &$F$\\
\hline
$\ \ \ -10\%\ \ \ \ \ \ \ \ \ \ \ \ \ \ \ \ \ \ \ \ \ \ \ \ \ $&$-10\%\ \ \ \ \ \ \ \ \ \ \ \ \ \ \ \ \ \ \ \ \ $&$-10\%\ \ \ \ \ \ \ \ \ \ \ \ \ \ \ \ \ \ \ \ $&$0.9469$\\
$10\%\ \ \ \ \ \ \ \ \ \ \ \ \ \ \ \ \ \ \ \ \ $&$-10\%\ \ \ \ \ \ \ \ \ \ \ \ \ \ \ \ \ \ \ \ \ $&$-10\%\ \ \ \ \ \ \ \ \ \ \ \ \ \ \ \ \ \ \ \ $&$0.9607$\\
$-10\%\ \ \ \ \ \ \ \ \ \ \ \ \ \ \ \ \ \ \ \ \ $&$10\%\ \ \ \ \ \ \ \ \ \ \ \ \ \ \ \ \ \ \ \ \ $&$-10\%\ \ \ \ \ \ \ \ \ \ \ \ \ \ \ \ \ \ \ \ $&$0.9853$\\
$-10\%\ \ \ \ \ \ \ \ \ \ \ \ \ \ \ \ \ \ \ \ \ $&$-10\%\ \ \ \ \ \ \ \ \ \ \ \ \ \ \ \ \ \ \ \ \ $&$10\%\ \ \ \ \ \ \ \ \ \ \ \ \ \ \ \ \ \ \ \ $&$0.9926$\\
$10\%\ \ \ \ \ \ \ \ \ \ \ \ \ \ \ \ \ \ \ \ \ $&$10\%\ \ \ \ \ \ \ \ \ \ \ \ \ \ \ \ \ \ \ \ \ $&$-10\%\ \ \ \ \ \ \ \ \ \ \ \ \ \ \ \ \ \ \ \ $&$0.9990$\\
$10\%\ \ \ \ \ \ \ \ \ \ \ \ \ \ \ \ \ \ \ \ \ $&$-10\%\ \ \ \ \ \ \ \ \ \ \ \ \ \ \ \ \ \ \ \ \ $&$ 10\%\ \ \ \ \ \ \ \ \ \ \ \ \ \ \ \ \ \ \ \ $&$0.9956$\\
$-10\%\ \ \ \ \ \ \ \ \ \ \ \ \ \ \ \ \ \ \ \ \ $&$10\%\ \ \ \ \ \ \ \ \ \ \ \ \ \ \ \ \ \ \ \ \ $&$10\%\ \ \ \ \ \ \ \ \ \ \ \ \ \ \ \ \ \ \ \ $&$0.9713$\\
$10\%\ \ \ \ \ \ \ \ \ \ \ \ \ \ \ \ \ \ \ \ \ $&$10\%\ \ \ \ \ \ \ \ \ \ \ \ \ \ \ \ \ \ \ \ \ $&$10\%\ \ \ \ \ \ \ \ \ \ \ \ \ \ \ \ \ \ \ \ $&$0.9531$\\
\hline \hline
\end{tabular} }
\end{center}

\begin{center}{\bf Table V. $\Omega_0T$ for STIRAP and corresponding fidelity $F$.\ \ \ \ \ \ \ \
\ \ \ \ \ \ \ \ \ \ \ \ \ \ \  }{\small
\begin{tabular}{cc}\hline\hline
$\Omega_0T$\ \ \ \ \ \ \ \ \ \ \ \ \ \ \ \ \ \ \ \ \ \ \ \ \ \ \ \ \ \ \ \ \ &$F$\\
\hline
$3.154$\ \ \ \ \ \ \ \ \ \ \ \ \ \ \ \ \ \ \ \ \ \ \ \ \ \ \ \ \ \ \ \ \ \ \ &$0.5538$\\
$5$\ \ \ \ \ \ \ \ \ \ \ \ \ \ \ \ \ \ \ \ \ \ \ \ \ \ \ \ \ \ \ \ \ \ \ &$0.6263$\\
$10$\ \ \ \ \ \ \ \ \ \ \ \ \ \ \ \ \ \ \ \ \ \ \ \ \ \ \ \ \ \ \ \ \ \ \ &$0.8516$\\
$15$\ \ \ \ \ \ \ \ \ \ \ \ \ \ \ \ \ \ \ \ \ \ \ \ \ \ \ \ \ \ \ \ \ \ \ &$0.9604$\\
$20$\ \ \ \ \ \ \ \ \ \ \ \ \ \ \ \ \ \ \ \ \ \ \ \ \ \ \ \ \ \ \ \ \ \ \ &$0.9898$\\
$25$\ \ \ \ \ \ \ \ \ \ \ \ \ \ \ \ \ \ \ \ \ \ \ \ \ \ \ \ \ \ \ \ \ \ \ &$0.9960$\\
$30$\ \ \ \ \ \ \ \ \ \ \ \ \ \ \ \ \ \ \ \ \ \ \ \ \ \ \ \ \ \ \ \ \ \ \ &$0.9992$\\
\hline \hline
\end{tabular} }
\end{center}

\newpage
\begin{figure}
 \centering
 \includegraphics[width=10cm]{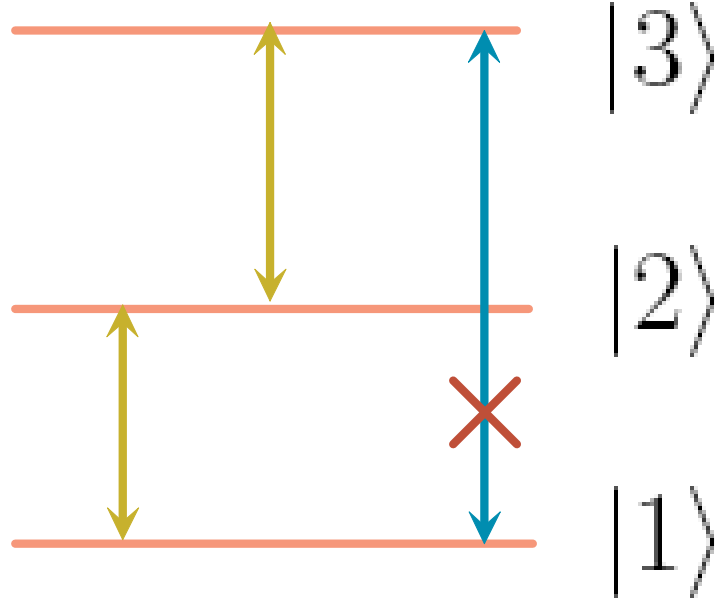}
 \caption{
         Energy levels of the three-energy-level Rydberg atom.
         }
 \label{fig1}
\end{figure}

\begin{figure}
 \centering
 \includegraphics[width=10cm]{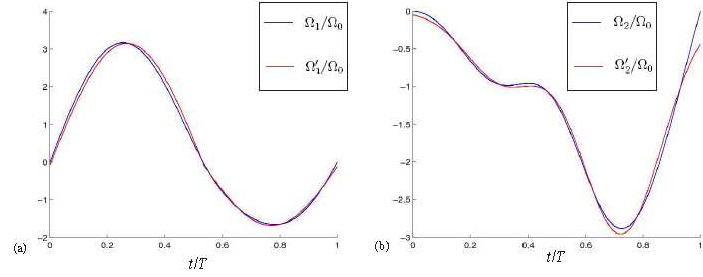}
 \caption{
         (a) $\Omega_1T$ and $\Omega_1'T$ versus $t/T$ with
$\mu=\pi/4$. (b) $\Omega_2T$ and $\Omega_2'T$ versus $t/T$ with
$\mu=\pi/4$ and $A=1$.
         }
 \label{fig2}
\end{figure}

\begin{figure}
 \centering
 \includegraphics[width=10cm]{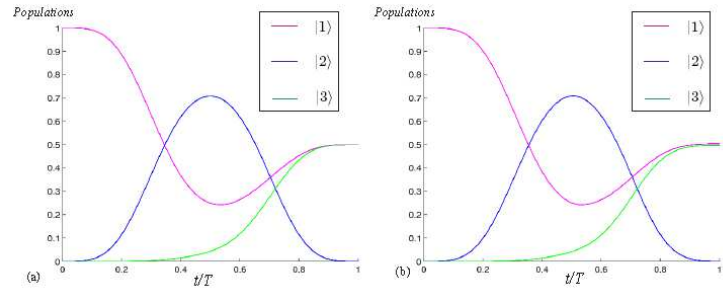}
 \caption{
         (a) Populations of states $|1\rangle$, $|2\rangle$ and
$|3\rangle$ versus $t/T$ when the Rydberg atom is driven by laser
pulses with Rabi frequencies $\Omega_1$ and $\Omega_2$. (b)
Populations of states $|1\rangle$, $|2\rangle$ and $|3\rangle$
versus $t/T$ when the Rydberg atom is driven by laser pulses with
Rabi frequencies $\Omega'_1$ and $\Omega'_2$. Here we set the
parameters $\mu=\pi/4$ and $A=1$.
         }
 \label{fig3}
\end{figure}

\begin{figure}
 \centering
 \includegraphics[width=10cm]{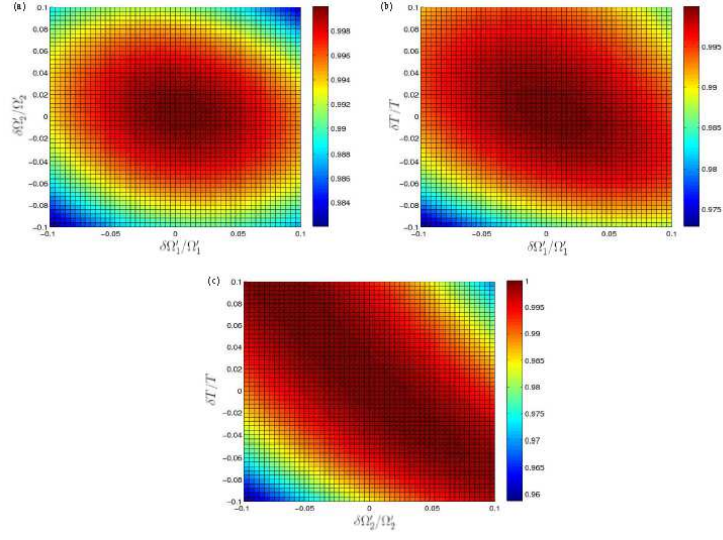}
 \caption{
         (a) Fidelity $F$ of the target state versus
$\delta\Omega_1'/\Omega_1'$ and $\delta\Omega_2'/\Omega_2'$. (b)
Fidelity $F$ of the target state versus $\delta\Omega_1'/\Omega_1'$
and $\delta T/T$. (c) Fidelity $F$ of the target state versus
$\delta\Omega_2'/\Omega_2'$ and $\delta T/T$. Here we set the
parameters $\mu=\pi/4$ and $A=1$.
         }
 \label{fig4}
\end{figure}

\begin{figure}
 \centering
 \includegraphics[width=10cm]{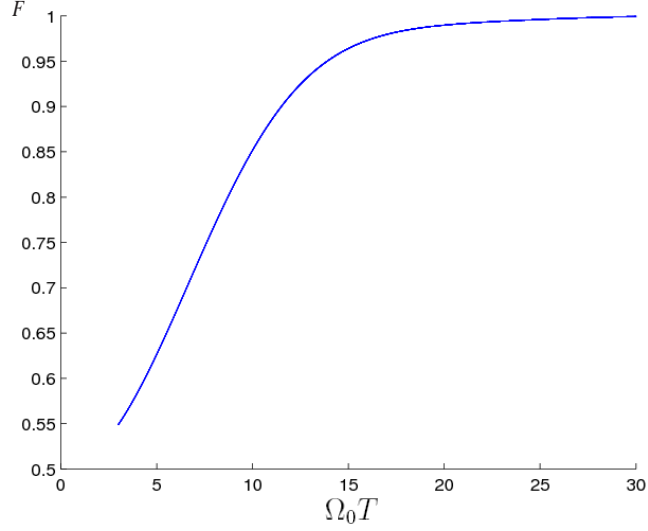}
 \caption{
         Fidelity $F$ of the target state versus $\Omega_0 T$ with
the STIRAP method.
         }
 \label{fig5}
\end{figure}

\begin{figure}
 \centering
 \includegraphics[width=10cm]{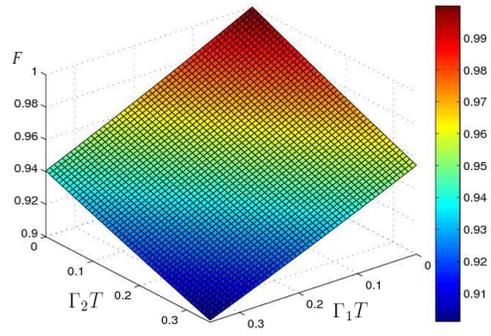}
 \caption{
         Fidelity $F$  of the target state versus $\Gamma_1/\Omega_0$
and $\Gamma_2/\Omega_0$.
         }
 \label{fig6}
\end{figure}

\end{document}